\documentclass[preprint,aps]{revtex4}

\usepackage{graphicx}
\usepackage{dcolumn}

\begin{document}

\preprint{Lebed-Bagmet}

\title{Quantum Limit in a Parallel Magnetic 
Field in Layered Conductors}

\author{A.G. Lebed$^{1,2}$ and N.N. Bagmet}

 \affiliation{$^{1}$Department of Physics, 
University of Arizona,
1118 E. 4-th Street, Tucson, AZ 
85721, USA}

\affiliation{$^{2}$Landau Institute for Theoretical Physics, 
2 Kosygina Street, Moscow, Russia}

\begin{abstract}
We show that electron wave functions 
 in a quasi-two-dimensional conductor 
 in a parallel magnetic field are always 
 localized on conducting 
 layers.
Wave functions and electron spectrum in a quantum limit, 
where the "sizes" of quasi-classical electron orbits are of 
the order of nano-scale distances between the layers,
are determined.
AC infrared measurements to investigate Fermi surfaces 
and to test Fermi liquid theory in Q2D organic and high-T$_c$ 
materials in high magnetic 
fields, $H \simeq 10 - 45 \ T$,
are suggested.
\end{abstract}

\maketitle
  
\pagebreak
  
Layered quasi-one-dimensional (Q1D) and  quasi-two-dimensional 
(Q2D) organic conductors exhibit unique magnetic 
properties [1-6]. 
Recently [4], it has been understood that most of them 
can be explained in terms of effective space dimensionality crossovers 
for electron wave functions in a magnetic 
field.
In their simplest forms, $3D \rightarrow 2D$ dimensional crossovers 
were suggested to explain field-induced spin-density wave phases 
[5-8,1] and to predict reentrant superconductivity (RS) 
phenomenon [8-10] .

In particular, in a Q2D conductor with electron spectrum,

\begin{equation}
\epsilon ({\bf p})= \epsilon_{\parallel}(p_x,p_y) 
+ 2t_{\perp}\cos(p_z  d) \  , \ \
\ t_{\perp} \ll \epsilon_{\parallel}(p_x,p_y) \sim \epsilon_{F} \ ,
\end{equation}
in a parallel magnetic field,
\begin{equation}
{\bf H} = (0,H,0) \  ,  \ \ \ \ {\bf A} = (-Hz,0,0) \ ,
\end{equation}
quasi-classical electron trajectories, determined by the equations
of motion,
\begin{equation}
d p_z / dt = e  v_x(p_x,p_y) H / c \  , 
\ \ v_x(p_x,p_y) = d \epsilon_{\parallel}(p_x,p_y) / d p_x \ ,
\end{equation}
become periodic and restricted along ${\bf z}$-axis:
\begin{eqnarray}
 &&z(t, p_x,p_y,H) = l_{\perp}(p_x,p_y,H) \cos[\omega_c(p_x,p_y,H) t] \ ,
\nonumber\\ 
 &&l_{\perp}(p_x, p_y,H) = 2 t_{\perp} / \omega_c (p_x,p_y,H) \ , \ \ 
\omega_c(p_x,p_y,H) = e v_x(p_x,p_y) H d / c \ .
\end{eqnarray} 
Periodic along ${\bf z}$-axis electron trajectories (4) correspond 
to "two-dimensionalization" of electron wave 
functions [9]. 
It is important that quasi-classical (QC) $3D \rightarrow 2D$ 
dimensional crossovers happen at weak magnetic fields [5,9], 
where the "sizes" of electron orbits (4) are much larger than 
the inter-plane distances, 
$l_{\perp}(p_x, p_y,H) \gg d$.
For instance, these QC crossovers are responsible for 
novel type of cyclotron resonance (CR) on open 
orbits [11-15].
In particular, Kovalev et al. [15] have suggested a new method 
to investigate Q2D Fermi surfaces (FS) by means of CR 
[11-14] and studied FS in organic conductor 
$\kappa$-(ET)$_2$I$_3$. 
For theoretical justification of the method, they used QC
kinetic equation, which is appropriate under experimental 
conditions [15], where $l_{\perp}(p_x, p_y,H) \geq d$  
at $H \simeq 1-5 \ T$ .

Meanwhile, in high experimental fields, $H \simeq 10-45 \ T$, 
typical "sizes" of electron orbits (4) become less than inter-layer 
distances [8], 
\begin{equation}
l_{\perp}(p_x, p_y,H) \leq d \simeq 10-30 \  A \ ,
\end{equation}
in a number of Q2D organic 
and high-T$_c$ materials.
Under condition (5) [which we call quantum limit (QL)], 
theoretical methods used so far [5-15] are not 
justified.
On the other hand, it is known that existence or not of 
Q2D Fermi surfaces
is one of the main problems in the area of high-T$_c$
and organic materials [16].
In this context, it is important to suggest quantum
mechanical variant of Kovalev et al. method [15] 
to investigate Q2D FS in high fields (i.e., in QL 
case (5)), where the method is less sensitive to impurities 
existing in doped high-T$_c$
materials.

The main goal of our Letter is to determine electron 
spectrum and wave functions in a Q2D conductor (1) 
in a parallel magnetic 
field (2).
We show that, in contrast to the extended Bloch waves 
[17,18], all wave functions are localized on conducting planes 
and are characterized by some quantum 
numbers  at $H \neq 0$.
Quantization law, obtained in this Letter, is qualitatively 
different from well-known Landau level quantization 
[17,18] in a perpendicular 
magnetic field.
As a result ac infrared properties are shown to be 
unusual.
In particular, we use our common results to extend QC 
method [15] to study Q2D FS  to QL
case (5).
We hope that this allows to test the existence of FS in
 numerous Q2D organic and 
 high-T$_c$ compounds.

To determine electron wave functions in Q2D conductor (1) 
in parallel magnetic field (2), we make use of QC 
description of electron motion within  conducting 
$({\bf x},{\bf y})$-planes and solve fully quantum mechanical 
problem for electron motion between the 
planes.
After QC Peierls substitutions for in-plane momenta, 
$p_x \rightarrow p_x - ( \frac{e}{c} ) A_x \ , 
p_x \rightarrow -i ( \frac{d}{dx} ) $, 
$p_y \rightarrow -i ( \frac{d}{dy} ) $ [17], 
one can represent electron Hamiltonian 
in the form:
 
\begin{equation}
 \hat{H} = \epsilon_{\parallel}\biggl( - i \frac{d}{dx} 
- \frac{eHz}{c},- i \frac{d}{dy} \biggl)+
\frac{1}{2m} \biggl( -i \frac{d}{dz} \biggl)^2 - 
\frac{V}{m} \sum^{ \infty }_{ n=-\infty } \delta (z- d n) \ ,
\end{equation}
where the last term introduces potential energy of crystalline 
lattice along ${\bf z}$-axis; $V >0$; $\delta(...)$ is 
Dirac delta-function.
Note that Hamiltonian (6) is exact one for an isotropic Q2D 
case.
 As it follows from general theory [17], the above mentioned 
 method disregards only corrections of the order of 
 $\omega^2_c(p_x,p_y,H) / \epsilon_F$ to electron energy
 for arbitrary  function 
 $\epsilon_{\parallel}(p_x,p_y)$. 
It is seen from Eqs. (4),(5) that QL condition corresponds 
to $ t_{\perp} \sim \omega_c(p_x,p_y,H)$, and, thus,  
these corrections  are of the order of
$ t^2_{\perp} / \epsilon_{F} \ll \omega_c (p_x,p_y,H)$,
where $\omega_c (p_x,p_y,H)$ is a characteristic energy 
scale in a
magnetic field.
Therefore, Hamiltonian (6) allows to study both QC and QL (5) 
dimensional crossovers. 

Arbitrary solution of the Schrodinger equation for Hamiltonian (6) 
can be written
as
\begin{equation}
\Psi_{\epsilon}(x,y,z)= \exp (ip_x x) \exp(i p_y y) \ 
\Psi_{\epsilon}(p_x,p_y;z) \ ,
\end{equation} 
which corresponds to free electron motion within 
$({\bf x},{\bf y})$-planes.
After substitution of Eq. (7) into Hamiltonian (6), 
it can be rewritten 
as follows:

\begin{equation}
 \hat{H} = \epsilon_{\parallel} \biggl(p_x - \frac{eHz}{c}, p_y \biggl) -
\biggl( \frac{1}{2m} \biggl)  \frac{d^2}{dz^2}  
 - \frac{V}{m} \sum^{ \infty }_{ n=-\infty } \delta (z- d n) \ .
\end{equation}
By expanding  in-plane energy in powers of $H$, 
it is easy to make sure that  Schrodinger equation for Hamiltonian 
(8) with the same accuracy can be expressed 
as: 
\begin{equation}
 \biggl[ - \biggl( \frac{1}{2m} \biggl) \frac{d^2}{dz^2} 
 -  \omega_c (p_x, p_y,H) \frac{z}{d}
- \frac{V}{m} \sum^{\infty}_{n=-\infty} \delta (z- d n) \biggl] 
 \Psi_{\epsilon}(p_x,p_y;z) =
 [\epsilon -  \epsilon_{\parallel} (p_x, p_y)] \Psi_{\epsilon}(p_x, p_y; z) \ .
\end{equation}
It is possible to prove [19] that, if one uses tight binding 
approximation for solutions of Eq.(9), 
\begin{equation}
 \Psi_{\epsilon}(p_x,p_y;z) = \sum^{\infty}_{m=-\infty} A_{m}(p_x,p_y) \ 
\Phi_{\epsilon_0}(z-d m)  \ 
\end{equation}
[where $\Phi_{\epsilon_0}(z-d m)$ is wave function of individual 
m-th layer at $H=0$, corresponding to energy 
$\epsilon_0 < 0$, $|\epsilon_0| \sim \epsilon_F$], then one 
disregards only corrections of the order of 
$ \omega^2_c(p_x,p_y,H) / [\epsilon_{\parallel}(p_x,p_y) , 
\epsilon_0] \sim t^2_{\perp} / \epsilon_F $ 
to electron
energy.

Therefore, equation
\begin{equation}
[\epsilon - \epsilon_0 - \epsilon_{\parallel} (p_x, p_y) 
+ m \omega_c (p_x, p_y,H)] 
A_{m}(p_x,p_y) = - A_{m+1}(p_x,p_y,H) t - A_{m-1}(p_x,p_y) t \ ,
\end{equation}
which can be derived after substitution of wave-functions (10) 
into Hamiltonian (9), has the same accuracy as Hamiltonian (6) 
and, thus, can be used to describe  $3D \rightarrow 2D$ 
QL dimensional 
crossovers (5).
At given in-plane momenta $p_x$ and $p_y$, Eq.(11) 
is equivalent to the so-called Stark-Wannier
 ladder equation in 
electric field [20]. 
Using Ref.[20], one can express wave functions and 
energy levels in the following way:
\begin{eqnarray}
&& \Psi_{N}(p_x,p_y;z) = \sum^{\infty}_{m=-\infty} 
J_{N-m}[2t_{\perp}/\omega_c( p_x,p_y,H)]  \  
\Phi_{\epsilon_0}(z-d m) \ ,
\nonumber\\
&&\epsilon_{N}(p_x,p_y) = 
\epsilon_{0} + \epsilon_{\parallel} (p_x,p_y) 
- N \omega_c(H, p_x,p_y) \ ,
\end{eqnarray}
where $J_{N}(...)$ is Bessel function of $N$-th 
order [21].
[An important difference between wave functions and energy 
spectrum (12) and that in Ref. [20] is that 
the envelope functions, $J_{N-m}(...)$, and energy 
levels, $\epsilon_{N}(...)$, in Eq.(12) depend on 
$p_x$ and $p_y$.]

Eq. (12) represents the main result of our Letter.  
In contrast to textbook extended Bloch waves with 
complex envelope, $\exp(i k z)$ [17], the envelope 
functions in Eq. (12) are real functions localized 
on the N-th conducting layer 
(see Fig.1).
Therefore, one concludes that, in a parallel magnetic field, 
all wave functions are localized on layers with energy gap  
between two neighboring wave functions being 
$\omega_c(p_x,p_y,H)$. 
Eq.(12) is valid both in QC and QL cases.

Below, we show that quantization law (12) 
leads to unusual ac infrared properties and suggest 
a method to investigate Q2D FS.
For these purposes, we calculate ac conductivity component, 
perpendicular to conducting layers, $\sigma_{\perp} (H, \omega)$, 
using known  wave functions  and energy 
spectrum (12).
Let us first find matrix elements of momentum  operator,
$\hat{p_z} = -i \frac{d}{dz}$, responsible for interactions between 
electrons and electric field, 
${\bf E} \parallel {\bf z}$.
It is possible to make sure that the matrix elements are non-zero 
only for wave functions with the same in-plane 
momenta, $p_x$ and $p_y$, and energies
$\epsilon_1 - \epsilon_2 = \pm \omega_c(p_x,p_y,H)$:
\begin{equation}
p_{z}^{N,N+1} = p_{z}^{N,N-1}  = 
 \int \Psi^*_{N} (z) \biggl( -i \frac{d}{dz} \biggl) \Psi_{N+1} (z) dz 
 = \int \Psi^*_{N} (z) \biggl( -i \frac{d}{dz} \biggl) \Psi_{N-1}(z)  dz  =
-i m d  t_{\perp}  \ .
\end{equation} 
[In other words, only optical transitions between electrons
with the same in-plane momenta localized on 
neighboring conducting layers are allowed].

To calculate $\sigma_{\perp} (H, \omega)$, we make use 
of the following extension [22] of Kubo 
formalism:
\begin{equation}
\sigma_{\perp}(H,\omega) = -i \frac{2 e^2}{m^2 V} 
\sum_{N_1,N_2} 
\frac{|p^{N_1,N_2}_z|^2}{(E_{N_1}-E_{N_2})} 
\frac{[n(E_{N_2})-n(E_{N_1})]}{( E_{N_2}-E_{N_1}-\omega-i \nu )} \ , 
\ \ \ \nu \rightarrow 0 \ ,
\end{equation} 
where $n(E)$ is Fermi-Dirac distribution function,
$V$ is a volume.
After substituting matrix elements (13) and energy
spectrum (12) in Eq.(14) and straightforward calculations, 
one obtains:
 \begin{equation}
\sigma_{\perp}(H,\omega) \sim
-i \ \int \ \frac{dp}{|{\bf v}_F (p_x,p_y)|} \ 
\biggl[ 
\frac{1}{\omega_c(p_x,p_y,H) - \omega -i \nu} 
+  \frac{1}{- \omega_c(p_x,p_y,H) - \omega -i \nu} 
\biggl]
\ , 
\ \ \ \nu \rightarrow 0 \  .
\end{equation} 
[Integration in Eq. (15) is made along 2D contour 
$\epsilon_{\parallel}(p_x,p_y) = \epsilon_F$;
${\bf v}_F(p_x,p_y) = d \epsilon_{\parallel}(p_x,p_y) /
d {\bf p}$; we use the approximation 
$n(E_{N_2}) - n(E_{N_1}) = (E_{N_2} - E_{N_1}) dn(E)/dE $ 
since $|E_{N_2} - E_{N_1}| = \omega_c(p_x,p_y,H)
 \ll \epsilon_F$ ].
 
It is convenient to write explicitly real and imaginary parts 
of conductivity (15):
 \begin{equation}
 \Re \ \sigma_{\perp}(H,\omega) \sim
 \int \ \frac{dp}{|{\bf v}_F (p_x,p_y)|} \  
 \biggl(
 \delta [\omega_c(p_x,p_y,H) - \omega] +
 \delta [\omega_c(p_x,p_y,H) + \omega] 
 \biggl) =
 \left\{
\begin{array}{c}
\neq 0 \ , \omega < \omega^{max}_c(H) \\
0 \ , \omega > \omega^{max}_c(H)
\end{array}\ \right .  
  \ ,
 \end{equation}
 \begin{equation}
\Im  \sigma_{\perp}(H,\omega) \sim
 \int \ \frac{dp}{|{\bf v}_F (p_x,p_y)|} \ 
\biggl[ 
\frac{1}{\omega_c(p_x,p_y,H) + \omega } 
-  \frac{1}{\omega_c(p_x,p_y,H) - \omega} 
\biggl]
\ ,
\end{equation} 
where $\omega^{max}_c(H)$ is the maximum value of energy 
gap, $\omega_c(p_x,p_y,H)$,  on the contour of 
integration (see Fig.2);
integral in Eq.(17) is determined as its principle 
value.

The main difference between Eqs.(16),(17) and the results
of Ref.[17] is that Eqs.(16),(17) are valid both in QC
and QL (5) cases, whereas the results [17] are essentially 
QC.
Another difference is that Eqs.(16),(17) describe
"optical" conductivity (i.e., conductivity in the absence of
impurities), in contrast to kinetic equation 
result [17].
From Eqs.(16),(17), it follows that ac properties in a 
parallel magnetic field 
are unusual.
Indeed, integration of $\delta$-function in Eq.(16) results in 
non-zero value of real part of conductivity  for ac frequencies 
 $0 < \omega < \omega^{max}_c(H)$
 (see Fig.2).
Therefore, electrons absorb electromagnetic waves 
 at $0 < \omega < \omega^{max}_c(H)$ (in the absence 
of impurities!), in contrast to text book properties 
of metals [18]. 

Let us demonstrate that real part of conductivity (16)
diverges at resonant frequency,
\begin{equation}
\omega = \omega^{max}_c (H) = e v^{max}_x(p_x,p_y) H d /c \ .
\end{equation} 
Indeed, in the vicinity of its maximum 
$\omega_c (p_x,p_y,H) \simeq
\omega^{max}_c (H) - A(H) |{\bf p}|^2$ with ${\bf p}$ being
momentum component perpendicular to ${\bf v}_F(p_x,p_y)$  
at point, where $|v_x(p_x,p_y)|$ takes its 
maximum
(see Fig.2).
In this case, integral (16)  can be estimated as
\begin{equation}
\Re \sigma_{\perp}(H,\omega) \sim 
\frac{1}{\sqrt{\omega^{max}_c-\omega}} \ , \ \ \ \ \
\omega^{max}_c(H) - \omega
\ll \omega^{max}_c(H) \ .
\end{equation} 
Therefore, by measuring $\omega^{max}_c(H)$
at different directions of the field one can determine 
angular dependence of $v^{max}_x(p_x,p_y,H)$
(see Ref.[15] and Fig.(2)).
We stress, however, that the physical meaning of
resonant frequency (18) at high magnetic fields (5),
where electrons are almost completely localized 
on conducting layers (see Figs.1,2),
is completely different from its kinetic equation
interpretation [15,23].

To summarize, wave functions and electron spectrum 
of a Q2D conductor in a parallel magnetic are 
determined.
A method to test FL picture in Q2D organic and 
high-T$_c$ materials is
suggested.
We hope that this method is a useful experimental tool 
to study Fermi-liquid versus non Fermi-liquid behavior
in low-dimensional compounds, especially as there 
have been claimed inconsistencies [24] between
angular resolved photo-emission methods [16] and
magneto-optical measurements in a perpendicular
magnetic 
field [24]. 
We also think that $3D \rightarrow 2D$ QL dimensional
crossovers and quantization law (12), suggested in the 
Letter, will be useful for studies of RS superconductivity 
[8-10] and for explanations of still unexplained phenomena 
observed in high parallel magnetic fields
[25,26].

One of us (AGL) is thankful to E.V. Brusse and P.M. Chaikin for
numerous and useful discussions.

\pagebreak

\begin{figure}[h]
\includegraphics[width=7.6in,clip]{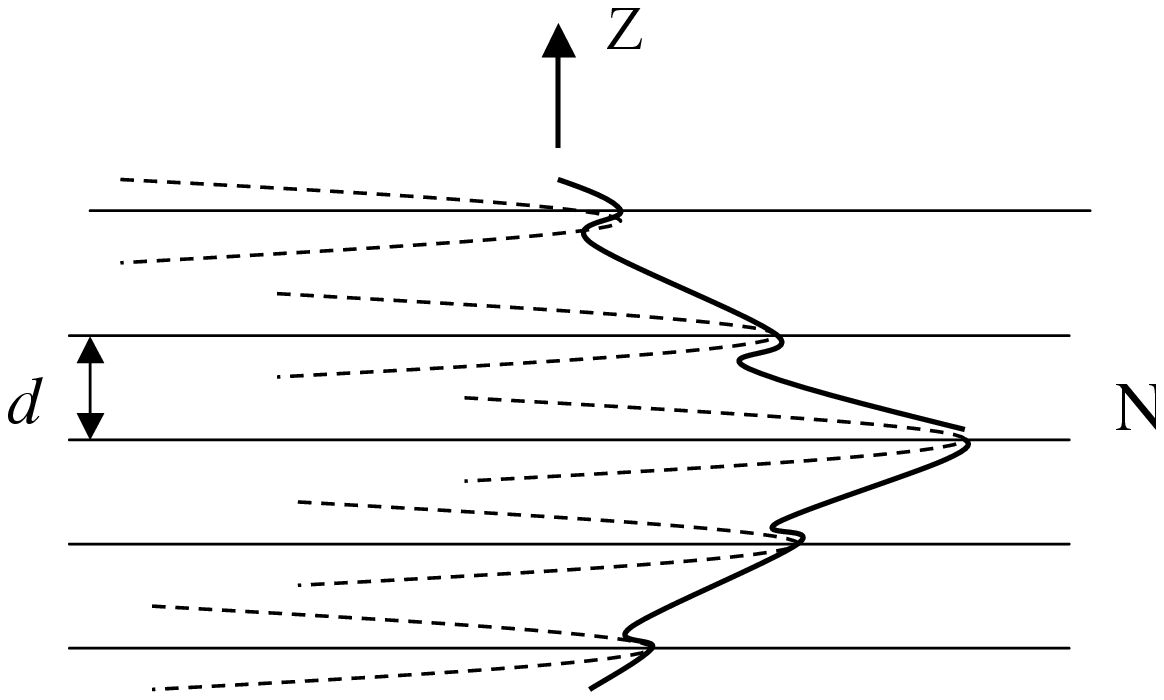}
\caption{ 
Wave function of a layered Q2D conductor (1) (at fixed values 
of $p_x$ and $p_y$), localized on $N$-th 
conducting layer in a parallel magnetic field (2),
is sketched.
Solid line: envelope function, $|J_{N-m}[2t_{\perp} / 
\omega_c(p_x,p_y,H)|$.
Dashed lines: wave functions of individual leyers, 
$\Phi_{\epsilon_0} (z - dm)$ (see Eqs.(10),(12)
and the text).
}
\label{fig1}
\end{figure}

\begin{figure}[h]
\includegraphics[width=7.5in,clip]{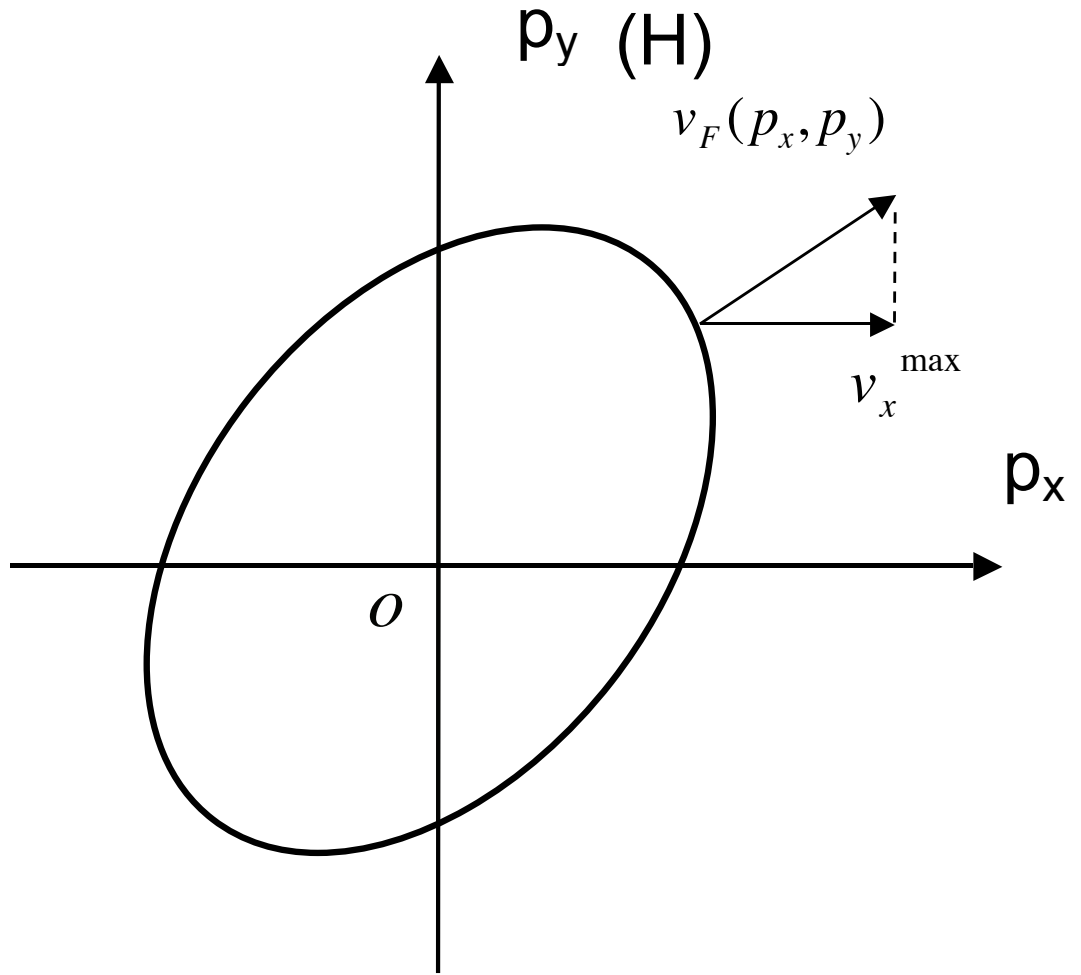}
\caption{ 
Resonant frequency $\omega^{\max}_c (H)$
corresponds to the maximum value of $|v_x(p_x,p_y)|$,
$v^{max}_x$,  on 2D Fermi surface, $\epsilon_{\parallel}(p_x,p_y) 
= \epsilon_F$, as it follows from Eq.(4).
}
\label{fig2}
\end{figure}

\end{document}